\documentclass{aa}  
\usepackage{hyperref}
\usepackage{float}

\usepackage{natbib}

\usepackage{graphicx}
\usepackage{txfonts,textcomp}

\begin{document}

\title{A complete census of planet-hosting binaries}

   \author{P. Thebault
          \inst{1}
          \and
          D. Bonanni\inst{2}
          }         
   \institute{LIRA, Observatoire de Paris, Universite PSL, 5 Place Jules Janssen, 92195 Meudon, France
       \and
       Sapienza Universita di Roma, Piazzale Aldo Moro 5, 00185 Rome, Italy
             }  
\offprints{P. Thebault} \mail{philippe.thebault@obspm.fr}
\date{Received ; accepted } \titlerunning{census of planet-hosting binaries}
\authorrunning{Thebault}

  \abstract
   {}
   {Estimating the effect binarity can have on planet-formation is of crucial importance, as almost half of field stars reside in multiple systems. One effective way to assess this effect is to get an accurate picture of the population of planet-hosting binaries and compare its characteristics to that of field star binaries.}    
   {We construct an extensive database, collected from intensive literature exploration, to achieve a complete census of all planet-hosting binaries known to date. Despite the heterogeneous character of the different surveys this database is built on, and the biases and selection effects that unavoidably affect any sample of planet-hosting binaries, we look for statistically significant trends and correlations within our sample}
   {Our database provides the characteristics (orbit or projected separation, stellar masses, distance, dynamical stability, etc.) for 759 systems (among which 31 circumbinaries), which is an increase by a factor of 9 with respect to the previous complete census of planet-hosting binaries. Of the 728 S-type systems, 651 are binaries, 73 are triples and 4 are quadruples. The raw distribution of planet-hosting binary separations peaks around 500\,au instead of 50\,au for field binaries. By analyzing the distribution of on-sky angular separations as a function of distance $d_b$ to the systems we argue that the observed deficit of planet-hosting close-in binaries cannot be explained solely by observational biases. Likewise, by exploring how multiplicity fractions among planet-hosts vary with $d_b$ we suggest that the subsample of known planet-hosting binaries at $<500\,$pc is not bias-dominated (but not bias-free). In this $<500\,$pc domain, the multiplicity fraction of planet-hosting stars is $\sim22.5\%$, approximately half of the value for field stars, and the deficit of planet-hosting binaries extends to separations of $\sim500\,$au, giving an approximate estimate of the detrimental effect binarity has on planet-formation}
   {}

   \maketitle

\section{Introduction}\label{sec:intro}

The multiplicity fraction of field stars in the solar neighborhood is almost 50$\%$ \citep{raghavan2010}. The study of exoplanets in the context of binaries is thus of crucial importance, given that binarity should be a relatively common environment. Theoretical studies have shown that some stages of the planet-formation process can be significantly affected by the presence of a stellar companion. Binary perturbations could, for example, truncate primordial protoplanetary disks \citep{savon94}, which could, on the one hand, deplete them of too much mass to form planets and, on the other hand, shorten their lifetime and may shorten disc lifetimes below that required for core accretion \citep[e.g.][]{muller12}. A stellar companion could also affect the planetesimal-accretion stage by increasing impact velocities to values that could prevent, or at least slow down growth by mutual accretion \citep{theb06,theb11}, even though disk self-gravity could counteract, to some extent, this detrimental effect \citep{sils21}. For a review on planet-formation in the context of binaries, see \cite{theb15} and \cite{marz19}.

Regardless of these theoretical considerations, another way to assess the influence of binarity on planet formation is to consider the population of observed exoplanet systems and look for potential differences between systems around single stars and systems in binaries. The most straightforward procedure would be to compare the occurrence rate of planets in single systems to that of planets in binaries. Such an analysis is, however, very difficult to perform in practice, as it requires precise information about the samples of binaries that have been explored in every planet-searching surveys. One of the few attempts at comparing planet occurrence rates in singles and binaries was carried out by \cite{hirsch21}, who considered a volume limited ($<$25pc) and unbiased sample of single and multiple stellar systems for which a relative velocity (RV) search for giant ($>0.1M_{\rm{Jup}}$) exoplanets was performed. They concluded that giant planet occurrence is $\sim12\%$ in binaries as compared to $\sim18\%$ for singles. However, the occurrence rate in wide $>100\,$au {(astronomical units) binaries seemed to be compatible with that in singles. It is only occurrences in close binaries that are significantly lower ($\sim4\%$), arguing for a detrimental effect of close binaries on planet formation. These results should, however, be considered with caution given the limited size of the considered binaries-with-planets sample (8 systems in total, and only one closer than 100au).

In practice, the most common way to investigate the effect binarity can have on planetary systems is to consider the problem the other way around, that is, to look for differences in the binarity fraction between exoplanet hosts and field stars. Early on, it has been noticed that the multiplicity fraction among planet hosts is relatively low: $23\%$ in \cite{ragha06}, $17\%$ in \cite{mugneu09}, or $12\%$ in \cite{roell12}, as compared to the derived value of $46\%$ for FGK field stars in the solar vicinity \citep{raghavan2010}. It was early on understood that these low values were, at least partially, due to strong observational biases. For the RV method, companion stars too close to the primary can indeed pollute the signal, so that binaries with separations $\lesssim$1-2" (and even sometimes up to 5") were initially excluded from most surveys \citep{eggen07,hirsch21}. The transit method does in principle not suffer from such strong adverse selection effects, even though it is not fully bias-free: the additional flux of the companion star decreases the transit depth and could make it sink below the (S/N) threshold for identifying, for example, Kepler Objects of Interest (KOIs) or Tess Objects of Interest (TOIs) \citep{wang15,ziegler20}. In addition, stars targeted by transit surveys, in particular the ones performed with the Kepler telescope, are generally relatively distant and had often not been vetted for binarity, so that, at least initially, a significant fraction could have been wrongly mislabeled as singles. To alleviate this issue, large surveys have been undertaken, using adaptive optics (AO) or "Lucky Imaging", to search for stellar companions around KOIs and TOIs \citep[e.g.,][]{kraus16,furlan17,ziegler20}. More recently this search for companions has been performed by mining the Gaia DR2 and DR3 catalogs to look for common parallaxes and proper motions \citep{mugmich20,mugmich21,fonta21,gonz24,michmug24}.

Interestingly, these surveys reach conclusions that sometimes contradict each other, even when considering the same category of planet-hosts. For the subsample of KOIs, for example, \cite{horch14} indeed conclude that the multiplicity fraction is comparable to that of field stars, while \cite{kraus16} find a strong deficit of planet-hosting binaries with separations $\lesssim$50-100au and \cite{wang14} conclude that the binary deficit even extends to separations of 1500au. We note, however, that the most recent surveys considering large samples of planet-hosts all find raw global multiplicity rates around 20$\%$, i.e., about one half of that for field stars : 23.2$\%$ in \cite{fonta21}, 19$\%$ for \cite{michmug24}, or 21.7$\%$ in \cite{gonz24}. There remains, however, a debate as to whether these low values are uniquely due to selection effects and the incompleteness of close (typically $\lesssim50$au)  binary censuses for distant targets (such as KOIs), or if this $\sim50\%$ deficit of planet-hosting binaries is real and physical: \cite{gonz24} attributes all the deficit to biases and adverse selection effects while \cite{fonta21} and \cite{michmug24} remain more cautious.

We present here a census of all known planet-hosting multiples to date,  compiling 759 systems (among which 31 circumbinaries), which corresponds to a factor 9 increase with respect to the last complete census of \cite{martin18}.
Our goal is twofold: firstly to provide a complete database to the community that will be regularly updated in an on-line accessible file, and secondly to use the unprecedented size of our sample (allowing us to explore correlations, such as multiplicity fraction as a function of distance, that had not been explored before) to derive statistically robust results about planet-hosting binaries characteristics despite the unavoidable observational biases affecting our sample.

The paper is organized as follows: Sec.\ref{sec:build} describes the procedure by which the database has been built and the information it provides. Sec.\ref{sec:results} presents the complete database and explores statistical correlations between crucial parameters (separation, system distance, multiplicity fractions, etc.). The robustness and reliability of our results is discussed in Sec.\ref{sec:discus}. Sec.\ref{sec:conclu} summarizes our results and main conclusions.

\section{Building the database} \label{sec:build}

\begin{table*}[ht!]
\begin{center}
\begin{tabular}{l l c l }
\hline\hline
Reference  & Method & Targets & Sample Size \\
\hline
\cite{patience02} &  AO and speckle imaging & 11 first RV hosts & 2 binaries 1 triple\\
\cite{ragha06} &  literature, photometry + astrometry & RV planet hosts & 30 binaries \\
\cite{mug06} &  imaging, photometry + astrometry &  & 3 binaries \\
\cite{eggen07,eggen11} &  AO imaging + astrometry & nearby stars & 4 binaries 2 triple\\
\cite{desid07} &  literature census&  & 37 multiple systems \\
\cite{bona07} &  literature & UD Sample & 15 multiple systems \\
\cite{mugneu09} & imaging + literature census &  & 43 multiple systems \\
\cite{howell11} & Speckle imaging & KOIs & 10 companion candidates\\
\cite{roell12} &  census from catalogs and literature &  & 57 multiple (15 new) \\
\cite{adams12,adams13} & AO imaging  & KOIs & 22 binary candidates \\
\cite{lillo12,lillo14} & lucky imaging & KOIs & 44 companion candidates\\
\cite{mug14} & lucky imaging + astrometry & Wide binaries & 7 multiple systems \\
\cite{wang14,wang15} & AO imaging, RV + color-analysis &KOIs & 59 candidates (6 bound)\\
\cite{horch14} & speckle imaging + stat. analysis &KOIs & 108 candidates (20 bound) \\
\cite{ngo15} & lucky imaging + astrometry & HJ hosts& 15 binaries and 2 triple \\
\cite{ginski16} & lucky imaging + astrometry & & 11 candidates (4 bound) \\
\cite{kraus16} & AO imaging & KOIs & 506 candidate binaries \\
\cite{ngo16} & lucky imaging + astrometry & HJ hosts& 27 stellar companions \\
\cite{evans16} & lucky imaging + astrometry & & 51 candidates (6 bound) \\
\cite{baranec16} & AO & KOIs & 203 candidate companions \\
\cite{ziegler17} & AO & KOIs & 221 candidate companions \\
\cite{ngo17} & lucky imaging + astrometry & RV giant hosts& 8 binaries and 3 triple \\
\cite{furlan17} & AO, speckle imaging and literature &87$\%$ of all KOIs& 2207 candidate companions \\
\cite{hirsch17} & color analysis of Furlan objects &$<$2" companions& 56 bound companions \\
\cite{martin18} &  literature census &  & 80 multiple systems \\
\cite{matson18} & speckle imaging & K2 hosts& 20 candidate companions \\
\cite{ziegler18} & AO + photometric vetting & KOIs& 42 bound companions \\
\cite{evans18} &  imaging, astro-/photometry, literature & transit  hosts & 7 bound companions \\
\cite{mug19} & Gaia DR2 astro- and photometry &$<$500pc& 176 binaries and 27 triple \\
\cite{ziegler20,ziegler21} & speckle imaging  & TOIs& 206 candidate multiples \\
\cite{mugmich20,mugmich21} & Gaia DR2 astro- and photometry & TOIs & 300 binaries and 20 triple  \\
\cite{hirsch21} & AO binary search, RV search for planets&$<$25pc& 8 planets in binaries \\
\cite{colton21} & speckle imaging + astrometry & KOIs& 31 binaries and 3 triple \\
\cite{fonta21} & Gaia DR2 and literature& $<$200pc &186 binaries and 32 triple  \\
\cite{michmug21} & Gaia DR2 astro- and photometry &$<$500pc& 41 binaries and 5 triple  \\
\cite{mug22,mugmich23} & Gaia DR2 astro- and photometry & TOIs & 258 binaries and 13 triple  \\
\cite{sull23,sull24} & AO + color analysis of Furlan KOIs& $<$2" companions & 326 binaries \\
\cite{gonz24} & Gaia DR3, WDS and literature & $<$100pc & 212 multiple-star systems\\
\cite{michmug24} & Gaia DR3 astro- and photometry &$<$625pc &311 binaries and 37 triple  \\
\cite{schlag24} & lucky imaging + astrometry &  & 33 binaries and 10 triple  \\
\cite{jing25} & neural networks on LAMOST spectra &  & 128 candidates  \\
\hline
\end{tabular}
\end{center}
\caption{Non-exhaustive list of main past surveys, looking for (or inventorying) stellar companions to exoplanet-hosts, which have been used in compiling our database of systems with "S-type" configurations. The "Sample Size" column  gives the number of planet-hosting binaries (and/or multiples) whose characteristics are listed in each study. This number can correspond to the systems effectively observed or analyzed in the survey \citep[as for example in][]{furlan17} or the number of already-known systems that have been compiled in the study \citep[for example in][]{roell12}, or a combination of the two \citep[as for example in][]{gonz24}. "Candidates" refers to stellar companions whose bound character could not be established (as for example in purely "lucky imaging" surveys).Note that the number of detected companions listed in the last column includes companions to hosts of planet candidates  }
\label{tab:studies}
\end{table*}

\subsection{Definition and boundaries}

Our database compiles all the \emph{confirmed} stellar companions to \emph{confirmed} (that is, not candidate) planet hosts. We adopt the conservative policy of excluding brown dwarfs and only considering hosts to "planets" having a mass (or a minimum mass) $\leq13M_{\rm{Jup}}$, the limit for Deuterium burning \citep{morl24}. Likewise, we only consider "stellar" companions of mass $M\geq75M_{\rm{Jup}}$, thus again excluding brown dwarfs \citep{chab23}. This allows us to have a well-defined separation between planetary and stellar companions. Our main database concerns systems with an "S-type" configuration, where the planet(s) orbits(-) one component of the binary, but we also compile a separate database for circumbinary cases ("P-type" orbits) that is presented in Sec.\ref{subsec:circum}

The following information is provided for all systems: name(s), masses of the stellar components, orbital parameters of the binary ($a_b$,$e_b$ and, for some rare cases, $i_b$), distance to the system, method of planet detection, number of known planets $n_{\rm{Pl}}$, mass ($m_{\rm{Pl}}$) and orbit ($a_{\rm{Pl}}$ and $e_{\rm{Pl}}$) of the outermost planet (i.e., the one most affected by the stellar companion's perturbations). 

\subsection{orbital stability}\label{sec:orbstab}

We also estimate, for each system, the outer limit for orbital stability $a_{\rm{crit}}$, which is the maximum radial distance to the planet-host star for which orbits are stable despite of the companion's perturbations. We derive $a_{\rm{crit}}$ using the widely-used empirical formula from \cite{holwig99}. Note that for most systems (especially those of separation $\geq\,50$au) only the projected separation between the stars is known while the actual orbit of the binary remains unconstrained. In this case the database only provides the projected separation $\rho_b$. To derive $a_{\rm{crit}}$ for these non-constrained orbits, we consider a "statistical" semi-major axis, using the average relation $log(a_b) = log(\rho_b) + 0.13$ derived by \cite{duq91} for randomly distributed orbits and a statistical eccentricity corresponding to the average value $<e_b>\sim0.45$ derived by \cite{raghavan2010} for field binaries near the sun. For these cases, the estimated stability limit thus only gives a first-order estimate and should be taken with caution. 

\subsection{higher-order multiples} \label{sec:hmul}

Some of the "binaries" listed in the database are in fact higher-order multiple systems, mostly triple or even quadruple stellar systems. However, most of these systems can be approximated as binaries from a dynamical point of view. The vast majority of them are indeed highly hierarchical and fall into two main categories:
\begin{enumerate}
    \item Systems where the companion star is itself a tight or spectroscopic binary. In this A-(BC) case, the dynamical behaviour of the circumprimary planetary system is very close to the case with one "virtual" stellar companion of mass $M_{\rm{B}}+M_{\rm{C}}$.
    \item Systems where the third star is much more distant (typically more than 10 times) from the central primary than the secondary. In this case the third star does not significantly impact the dynamical evolution of the planetary system and its influence can be neglected in a first approximation
\end{enumerate}
This is why, for the sake of clarity and simplicity, and in order not to artificially reduce the size of the sample on which we derive general trends (see Sec.\ref{sec:results}), those hierarchical triples are presented as binaries in the main database, with either a "virtual" merged BC companion or neglecting a too distant C component. The dynamical stability outer limit ($a_{\rm{crit}}$) of the planet is then computed with these simplified approximations. These hierarchical triple are labeled with an added "*" suffix at the end of the system's name. The exact description of these systems is given in the "Notes and References on individual systems" Table \ref{tab:notes}

\subsection{Sources}

Our database is mostly based on a deep and thorough literature search. Our main sources are surveys looking for stellar companions to planet-hosting stars. As mentioned in Sec.\ref{sec:intro}, these surveys are mainly of two different types: high-precision imaging (AO, speckle) or mining of the Gaia DR2 and DR3 database. Note that we only compile \emph{confirmed} stellar companions, so that companions identified in imaging surveys are considered only if there is an additional analysis that can distinguish between a physical bound companion and a background object. This additional vetting is generally either astrometric (looking for common parallaxes and proper motions) or photometric (multi-band analysis to check if colors are consistent with a main-sequence companion rather than a background object). We make an exception to this rule for a handful of companions detected in imagery for which the statistical chances for being a background object are less than $1\%$ \citep{ziegler20}. Note that we also only considered systems with \emph{confirmed} (i.e., not candidate) planets, thus eliminating a significant fraction of stellar companions found around KOIs or TOIs that have not yet been confirmed as planets. It is necessary to regularly go back to these surveys as some planets that were candidates by the time of the survey later become confirmed.

In some cases, the binarity of the system is not assessed in surveys but directly in the planet-discovery study. This is notably the case for most systems detected by microlensing \citep[e.g.][]{gould14,poleski14}, but also for a non-negligible number of RV or transit detections \citep[e.g.][]{howard10,buch11,beatty12}. Finally, there is a small (but non-negligible) fraction of our sample for which the binarity has been long known and taken from catalogs such as the Washington Double Star Catalog  \citep[WDS,][]{mason01}.

There is a significant fraction of stellar companions that have been identified in several separate surveys. In these cases, we try to give credit (in the "Notes and References on individual systems" Tab.\ref{tab:notes}) to the earliest work having assessed the binarity. With respect to the binary characteristics (stellar mass, distance, projected separation, etc.), we consider the constraints obtained from the most recent studies.

Tab.\ref{tab:studies} gives a non-exhaustive list of the main sources that we have used.

\section{Results} \label{sec:results}

\begin{table*}[ht!]
\begin{center}
\begin{tabular}{c c c c c c c c c c c c c c}
\hline
Name  & Alt. Name &$M_1$ & $M_2$ & $d$ & method & $a_b$ (or $\rho_b$) & $e_b$ & $n_{\rm{Pl}}$ & $a_{\rm{Pl}}$ & $e_{\rm{Pl}}$ & $m_{\rm{Pl}}$ or $m.sini$& $a_{\rm{crit}}$ & $i_{\rm{b}}$\\
 & (if any) & ($M_{\oplus}$) & ($M_{\oplus}$) & (pc) & & (au) & & & (au) & & ($M_{\rm{Jup}}$) & (au) & (deg.)\\
\hline\hline
HD42936 & DMPP-3 & 0.900 & 0.087 & 48.90 & 2 & 1.220 & 0.594 & 1 & 0.066 & 0.14 & 0.0077 & 0.16 & - \\
Kepler693 & KOI824 & 0.760 & 0.150 & 1002. & 1 & 2.900 & 0.480 & 1 & 0.112 & - & 0.800 & 0.50 & 53.0 \\
TOI179 & HD18599 & 0.807 & 0.080 & 38.58 & 1 & 3.520 & - & 1 & 0.048 & 0.34 & 0.0758 & 0.93 & -\\ 
Kepler420 & KOI1257 & 0.990 & 0.700 & 900. & 1 & 5.300 & 0.310 & 1 & 0.382 & 0.77 & 1.45 & 1.02 & 18.\\ 
Kepler968 & KOI1833 & 0.760 & 0.390 & 312. & 1 & 5.300 & - & 3 & 0.063 & 0. & 0.019 & 1.13 & -\\ 
Kepler1649 & KOI3138 & 0.310 & 0.190 & 92.4 & 1 & 5.540 & - & 2 & 0.096 & - & 0.0025 & 1.13 & -\\ 
HIP90988 & HD170707 & 1.300 & 0.097 & 93.18 & 2 & 7.480 & 0.289 & 1 & 1.36 & 0.08 & 2.10 & 2.08 & -\\ 
HD72892 & - & 1.020 & 0.075 & 72.78 & 2 & 8.124 & 0.375 & 1 & 0.228 & 0.42 & 5.45 & 1.92& -\\ 
TOI1736 & - & 1.000 & 0.780 & 88.90 & 1 & 8.200 & - & 2 & 1.35 & 0.16 & 8.26 & 1.57 & -\\ 
HD144899 & - & 1.160 & 0.095 & 113.0 & 2 & 8.332 & 0.355 & 1 & 0.242 & 0.82 & 0.064 & 2.03 & -\\ 
HD28192 & - & 1.080 & 0.090 & 49.80 & 2 & 8.500 & 0.029 & 1 & 0.118 & 0.10 & 0.314 & 3.55 & -\\ 
HIP86221* & - & 0.790 & 0.510 & 29.48 & 2 & 8.500 & 0.205 & 1 & 0.031 & 0.09 & 0.710 & 2.00 & 165\\ 
........\\
\hline
\end{tabular}
\end{center}
\caption{Complete database of planet-hosting "S-type" binaries, sorted by increasing binary separation. This is just
an excerpt of the full table that is available, in a machine-readable form, at this url: \url{https://exoplanet.eu/planets_binary/} }
\label{tab:database}
\end{table*}

Table \ref{tab:database} presents our complete database, which compiles 728 "S-type" planet-hosting multiples. This is more than twice the size of the largest samples considered in earlier studies \citep[348 systems in][]{michmug24} and almost 10 times the size of the last published full census of planets in binaries \citep{martin18}.
This database will be regularly updated and will be available to the community in an online version  (\url{https://exoplanet.eu/planets_binary/}), hosted on the  "encyclopedia of exoplanetary systems" website \citep{schn11}.

\subsection{separation distribution}\label{sec:sepdis}

We present in Fig.\ref{fig:sep} a histogram of the distribution of binary semi-major axis $a_b$. This distribution peaks at around 500\,au, which corresponds to a pronounced deficit of close-in companions when compared to the canonical distribution of \cite{raghavan2010}. 
Using a bootstrap resampling procedure on the $log(a_b)$ distribution of our planet-hosting binaries sample we get $<a_b>=737^{+133}_{-88}$\,au at the $3\sigma$ confidence level, confirming that this distribution is statistically different from that of field binaries which peaks at 45\,au \footnote{ Histograms produced with different prescriptions between projected separation $\rho_b$ and real semi-major axis $a_b$ do not lead to statistically different results. As an example, the extreme case $a_b=\rho_b$ leads to $<a_b>=686^{+117}_{-79}$\,au}.
This deficit of tight planet-hosting binaries is a well known feature that had been identified in earlier studies for smaller binary samples \citep[e.g.,][]{kraus16}. As mentioned in Sec.\ref{sec:intro}, the main question is to what extent this depletion is due to observational biases or adverse selection effects and to what extent it is a real and physical feature. 
Answering that question might appear less straightforward for the present sample, given the all-encompassing nature of our census and the heterogeneity of the sources it is built upon, than for more limited and homogeneous surveys (such as volume-limited studies targeting only KOIs or TOIs). The size of our sample does, however, allow us to explore parameters that can help put constraints on the level of bias (or lack thereof) affecting our census. 

A compelling indication that the lack of planet-hosting tight binaries cannot be entirely due to observational biases follows from plotting binary projected \emph{angular} separations ($\theta_b$) as a function of distance $d_b$ to the system (Fig.\ref{fig:distang}). Since the main adverse selection effects (see Sec.\ref{sec:intro}) against binaries are, to a first order, inversely proportional to the angular separation between stellar components, we would indeed expect the lower envelope of the ($\theta_b$,$d_b$) distribution to be horizontal in a bias-dominated sample. This is clearly not what is observed in Fig.\ref{fig:distang}, which shows that, on the contrary, the lower envelope of the ($\theta_b$,$d_b$) distribution follows an inclined $\rho_b\sim$cst line (see Sec.\ref{sec:closein} for more discussion on this issue).

\begin{figure}[h!]
\includegraphics[scale=0.5]{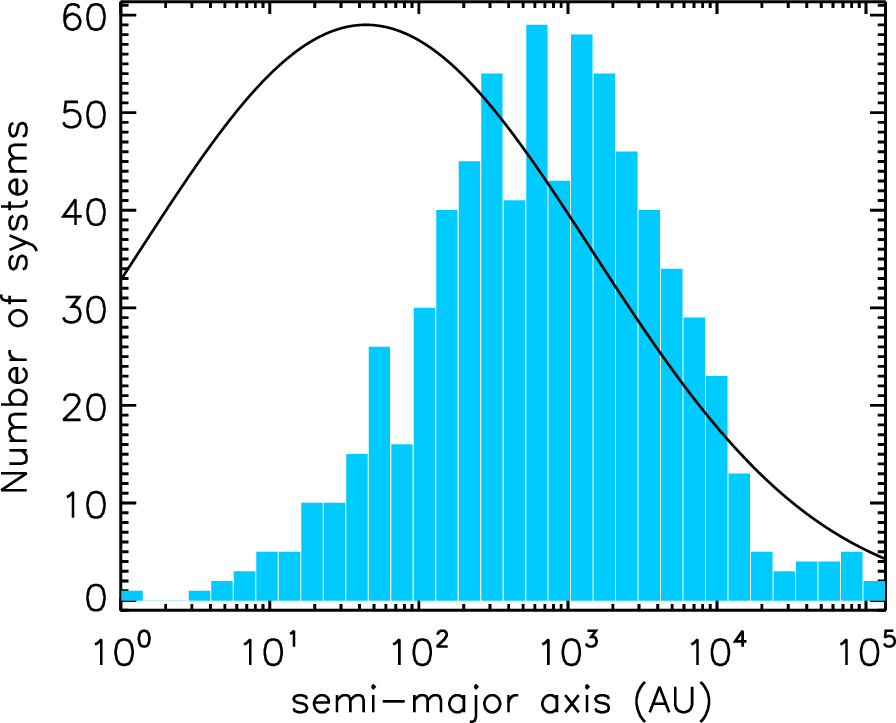}
\caption{{Histogram of planet-hosting binary semi-major axis for the complete sample. When only the projected separation $\rho_b$ is known, a "statistical" semi-major axis is considered following the relation derived by \cite{duq91}} for randomly distributed orbits (see text for details). The black line corresponds to the normalized distribution for field stars derived by \cite{raghavan2010}.
\label{fig:sep}}
\end{figure}

\begin{figure}[h!]
\includegraphics[scale=0.5]{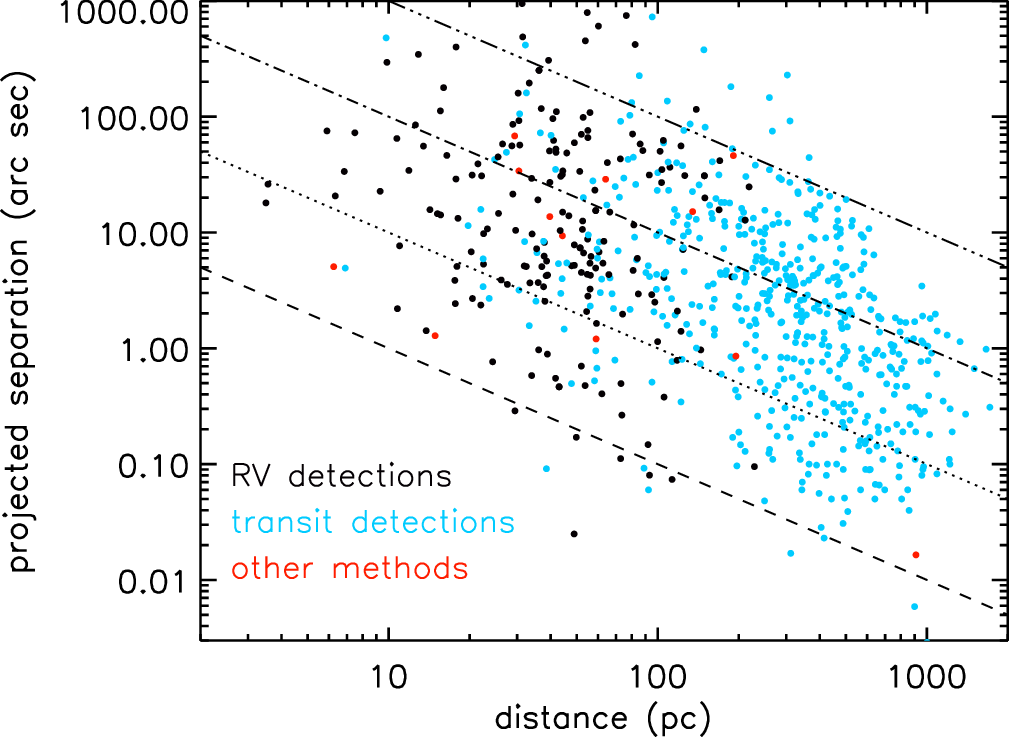}
\caption{{On-sky projected angular separation $\theta_b$ between binary components as a function of distance $d_b$ to the system. The 4 black lines correspond (from bottom to top) to physical separations of 10, 100, 1000 and 10000au. Note that Proxima Centauri has been left out of this graph in order not to squeeze the graph's Y-axis.}.
\label{fig:distang}}
\end{figure}

\subsection{multiplicity fraction}\label{sec:mulfra}

\begin{table}[h!]
\begin{center}
multiplicity fraction
\begin{tabular}{c r r c}
\hline
  & all distances & $\leq\,$500pc & field stars\\
\hline\hline
full sample & $16.1\%$ & $22.5\%$ & $46\%$ \\
\hline
M primaries & $9.2\%$ & $11.8\%$  & $35\%$ \\
K primaries & $15.2\%$ & $19.1\%$ & $41\%$ \\
G primaries & $14.6\%$ & $24.0\%$ & $45\%$ \\
F primaries & $22.9\%$ & $33.7\%$ & $51\%$  \\
A primaries & $23.7\%$ & $26.7\%$ & $69\%$  \\
\hline
\end{tabular}
\end{center}
\caption{Multiplicity fraction $f_M$ for the whole population of exoplanet-hosts, as well as as a function of stellar type (for the planet-bearing star). We show, as a reference, the $f_M$ estimated by \cite{raghavan2010} for field stars in the solar neighborhood.}
\label{tab:mulfra}
\end{table}

\begin{figure}[h!]
\includegraphics[scale=0.5]{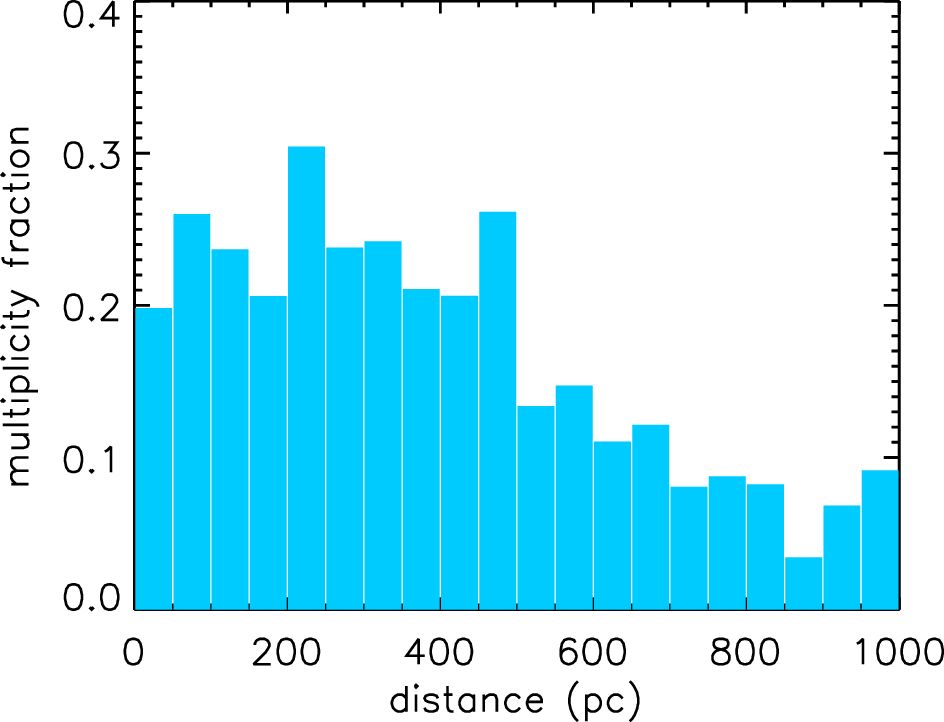}
\caption{{Multiplicity fraction of confirmed planet-hosts as a function of distance to the system}.
\label{fig:distfra}}
\end{figure}

To compute multiplicity fractions amongst planet host stars, we take as a reference the up-to-date population of planet-hosting systems in the encyclopedia of exoplanetary systems (\url{https://exoplanet.eu}). As of May 2025, the total number of systems hosting at least one $\leq13M_{\rm{Jup}}$ planet is 4504. This gives a raw multiplicity fraction $f_M$ for the whole sample that is equal to 728/4504=0.161. This is slightly lower than the $\sim20\%$ found in recent surveys (see Sec.\ref{sec:intro}) and much lower than the 0.46 fraction derived by \cite{raghavan2010} for field stars.

As with the separation distribution, the issue is here to assess to what extent this $f_M$ value is affected by observation biases and selection effects. And, as for the separation distribution, plotting how this parameter varies with distance can provide some important clues. As can be seen on Fig.\ref{fig:distfra}, the multiplicity fraction is indeed almost constant in the $d_b<500\,$pc domain, which is not what would be expected for a sample dominated by biases, for which $f_M$ should decrease with increasing distance. This is here again because the main adverse biases affecting our sample become more pronounced with decreasing on-sky \emph{angular} separation between the binary components, thus leading to lower $f_M$ values at larger $d_b$. This expected decrease of $f_M$ values is in fact observed, but only beyond $\sim500\,$pc \footnote{ The difference between $f_M$ values in the $d\leq500\,$au and $d\geq500\,$au domains is statistically significant, with the $3\sigma$ error bars being $<f_M>_{d<500}=0.225^{+0.013}_{-0.012}$ and $<f_M>_{d>500}=0.120^{+0.027}_{-0.025}$} , indicating that it is only in this distance domain that the sample becomes bias-dominated.
We discuss these issues in more detail in Sec.\ref{sec:mul500}, but we will assume from here on that the region at $d_b<500\,$pc provides global statistics that are, to a first order, reliable. For these $d_b<500\,$pc systems, the global multiplicity fraction increases to $f_M=22.5\%$ (Tab.\ref{tab:mulfra}), which is comparable to the 23.2$\%$ in \cite{fonta21}, 19$\%$ for \cite{michmug24}, or 21.7$\%$ in \cite{gonz24}. We argue that this value is probably close to the real unbiased value for planet-bearing binaries (see discussion in Sec.\ref{sec:discus}).

\subsection{stellar types}

\begin{figure}[h!]
\includegraphics[scale=0.5]{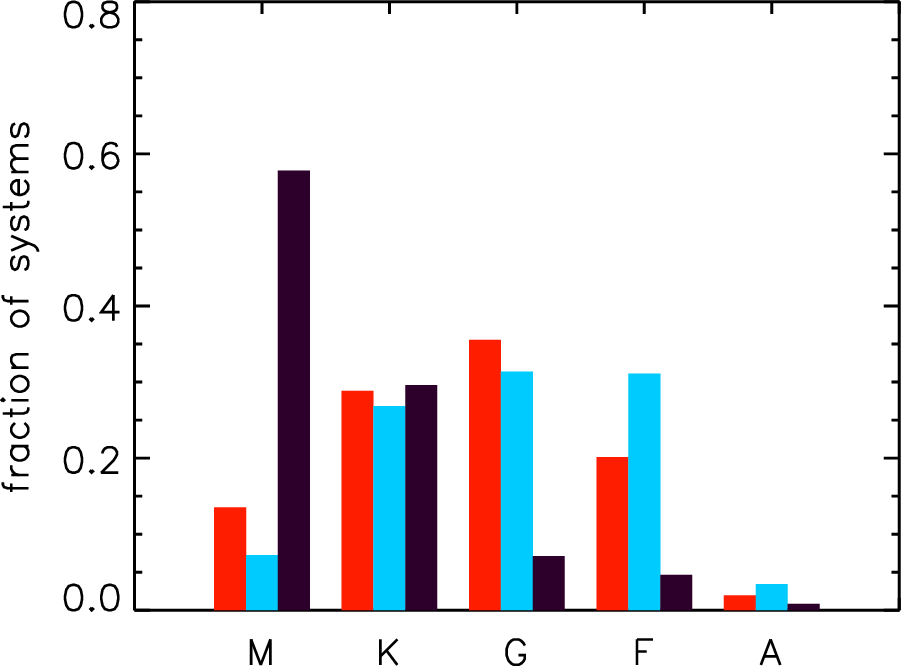}
\caption{{Fraction of planet-hosting primaries (blue) and stellar companions (deep purple) in our database as a function of stellar type. The red histogram corresponds to the distribution for planet-hosting single stars (taken from the encyclopedia of exoplanetary systems).}
\label{fig:stype}}
\end{figure}

The vast majority of the planet-hosting primaries consists of K-G-F stellar types (Fig.\ref{fig:stype}), which make up more than 90$\%$ of our sample. We note that this stellar-type distribution of primaries is a close match to that for planet-hosting single stars, with a slight deficit of M stars and a slight excess of F types.
The population of companion stars is, on the other hand, completely dominated by low-mass M-stars, which account for $58\%$ of this subsample. These results agree relatively well with those obtained by \cite{fonta21} on a more limited sample.

When looking at the multiplicity fraction for each stellar type (Tab.\ref{tab:mulfra}) we see that it progressively increases from M to F stars, which does qualitatively agree with the trend observed among field binaries \citep{raghavan2010}. The differences between different stellar types are, however, more pronounced than for field binaries. As an example, $f_M(F_{\rm{stars}})/f_M(M_{\rm{stars}})\sim3$ for planet-hosts whereas this ratio is only equal to $\sim1.6$ for field stars. This could indicate that the detrimental effect of binarity on planet formation increases with decreasing stellar mass.

\subsection{triples and higher-order multiples}

\subsubsection{occurrence rate}

Our census lists 77 higher-order multiples (73 triples and 4 quadruples), corresponding to $10.6\%$ (or $12.4\%$ if only considering systems at $<500$pc) of our total sample of 728 planet-hosting S-type multiples. This value is close to the one obtained by \cite{michmug24}, $12\%$, but is significantly lower than the $19.5\%$ for field stars found by \cite{raghavan2010}. As pointed out by \cite{gonz24}, such lower values might be due to the fact that, if planet-hosting primaries have been investigated by companion-search surveys, their identified companions (especially wide ones) have not been investigated as thoroughly, so that a significant fraction of wide planet-hosting binaries might in fact be triples (the companion star being itself a yet undetected binary).

\subsubsection{non-hierarchical triples}

As mentioned in Sec.\ref{sec:hmul}, the vast majority of our subsample of triples is hierarchical, with a ratio between the separations of the two different binary pairs that is higher than 10. We identify, however, 3 rare non-hierarchical triples (see Tab.\ref{tab:nonhier}), for which the dynamical role of the third component cannot in principle be neglected. We note, however, that for all these 3 systems only the projected separations $\rho_{\rm{AB}}$ and $\rho_{\rm{AC}}$ of the stellar companions are known and the inclination between the A-B and B-C orbital planes is not constrained. We can thus not rule out that these 3 cases are in fact hierarchical because the third component is in reality much further away from the primary than $\rho_{\rm{AC}}$.

\begin{table}[h!]
\begin{center}
\begin{tabular}{c c c c }
& $a_{\rm{Pl}}$ & $\rho_{\rm{AB}}$ & $\rho_{\rm{AC}}$\\
& (au) & (au) & (au) \\ 
  \hline\hline
HD213885 & 0.057 & 23.7 & 55.1\\
HD207496 & 0.063 & 63.7 & 179.6 \\
Kepler-25 & 0.112 & 2030 & 7771 \\
\hline
\hline
\end{tabular}
\end{center}
\caption{Planet semi-major axis ($a_{\rm{Pl}}$) and stellar companions projected separations ($\rho_{\rm{AB}}$ and $\rho_{\rm{AC}}$) for the 3 non-hierarchical planet-bearing triples. }
\label{tab:nonhier}
\end{table}

\subsubsection{Planets that are both on S-and P-type orbits}

For four systems, HD155555, BX Tri, Kepler-64 and KIC7177553, it is the central planet-hosting star that is itself a very tight binary ($a_b\leq0.2\,$au). In these  cases, the planet is actually on \emph{both} an S-type \emph{and} a P-type orbit. Since all four systems are highly hierarchical we list them both in our main database as S-types, for which we assume a "virtual" primary by adding the masses of the 2 inner components, and in our separate database of circumbinary systems (see Sec.\ref{subsec:circum}), in which we neglect the influence of the third outer stellar component.

\subsection{Architecture and stability}

\begin{figure*}[h!]
\includegraphics[scale=0.86]{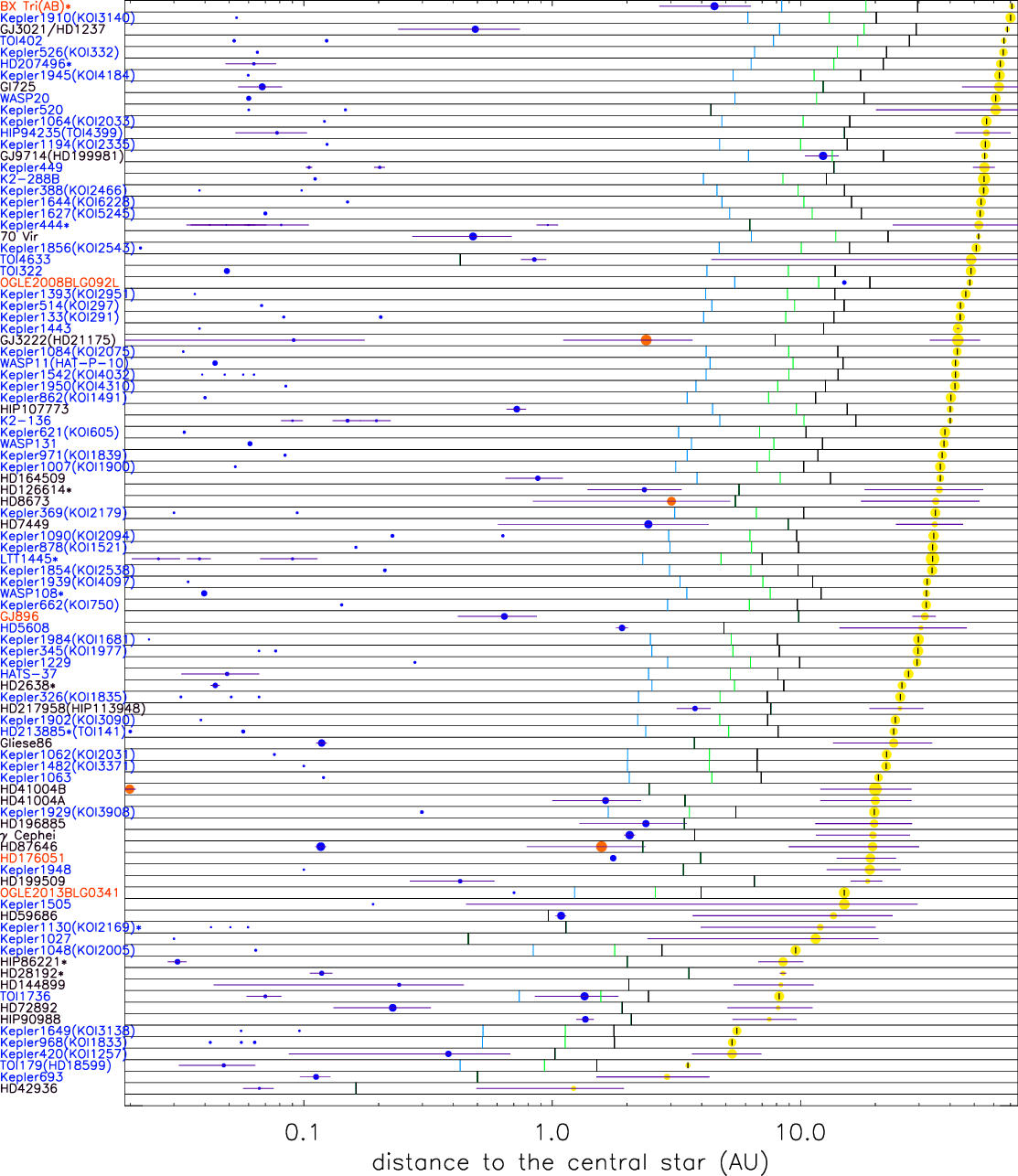}
\caption{{Architecture of the 100 tightest planet-hosting binaries. Blue circles show $a_{\rm{Pl}}$ while yellow circles show $a_b$. The radius of the blue circle is proportional to the estimated radius of the planet. The radius of the yellow circle is proportional to the cubic root of the mass ratio between the companion and the central star (for the sake of visibility, planet sizes have been inflated by a factor 20 with respect to star sizes). For both planets and stars, the horizontal purple line represents the radial excursion due to the object's eccentricity. For systems for which the semi-major axis of the binary is unknown, we simply plot the projected separation between the stellar components. These cases are indicated by a "$|$" symbol overlaid onto the companion star. The black vertical bars between the planet and the stellar companion show the outer limit $a_{\rm{crit}}$ for long-term orbital stability as estimated with the empirical formula of \cite{holwig99}. When only the projected separation $\rho_b$ of the binary is known, then the black bar represents $a_{\rm{crit}}$ for a system with $a_b=\rho_b$ and $e_b=0$, while an additional green bar shows $a_{\rm{crit}}$ assuming a "statistical" binary orbit ($log(a_b) = log(\rho_b) + 0.13$ and $e_b=0.45$) and an additional blue bar represents a "high-eccentricity" case with $a_b = \rho_b$ and $e_b=0.7$ . Systems detected by the RV method are written in black, those detected by transit are in blue, and those detected by other methods are in red. For systems where a brown dwarf exists \emph{in addition} to planet(s), the BD location is plotted in orange}.
\label{fig:binstruc}}
\end{figure*}

Fig.\ref{fig:binstruc} presents the dynamical architecture for the 100 tightest planet-hosting binaries, with projected separation up to $\sim75\,$au. 
For each system we plot (as a vertical bar) the location of the outer planetary orbit stability limit $a_{\rm{crit}}$ given, as a function of $a_b$, $e_b$ and $\mu_b$ (mass ratio between the binary components), by the empirical formula of \cite{holwig99}. For the cases where $a_b$ and $e_b$ are unconstrained and only the projected separation $\rho_b$ is known, we consider three different ($a_b$, $e_b$) configurations (and thus three different $a_{\rm{crit}}$): one reference case assuming the "statistical orbit" $log(a_b) = log(\rho_b) + 0.13$ and $e_b=0.45$ (see Sec.\ref{sec:orbstab}), one "low $e_b$" case with $a_b=\rho_b$ and $e_b=0$ and one "high-$e_b$" case with $a_b=\rho_b$ and $e_b=0.7$. 
It can be seen that, for almost all systems, planets are located in the dynamically "safe" $a_p\leq a_{\rm{crit}}$ region around the planet-hosting star. There are only 2 systems with constrained binary orbits for which $a_p$ lies beyond $a_{\rm{crit}}$: HD 59686 and TOI-4633. In both cases it is the very high value of $e_b$, 0.91 for TOI-4633 and 0.73 for HD 59686, which is responsible for this dynamically "unsafe" situation. \cite{trifo18} thoroughly investigated HD 59686's dynamics and found that there is nevertheless a relatively large parameter space for dynamical stability for coplanar but retrograde orbits \citep[a configuration that was not considered in][]{holwig99}. 

For systems with unconstrained binary orbits, there is no case for which planets are located beyond all three different $a_{\rm{crit}}$. For one system, OGLE-2008BLG-092L, the $a_p$ of the outermost planet is beyond both the "statistical" and the "high-$e_b$" $a_{\rm{crit}}$, but this result should be taken with caution given that the mass of the companion is largely unconstrained \citep[it could even be a brown dwarf, see][]{poleski14}. It remains, nevertheless, the system with the highest $a_{\rm{p}}/\rho_b$ ratio (close to 1/3). Finally, there are two systems, TOI-1736 and GJ 9714, for which the $a_p$ of the outermost planet is beyond the "high-$e_b$" $a_{\rm{crit}}$ 

We stress that the $a_b\leq a_{\rm{crit}}$ criteria of \cite{holwig99} should only be taken as a first order estimate. It is for instance only valid for coplanar orbits and breaks down for $e_b\geq0.7$-$0.8$. It also ignores complex islands of stability such as mean motion resonances \citep{pilat05}. Last but not least, let us repeat that, in many cases, the orbital parameters (notably $a_b$ and $e_b$) of the binary are not known, and that the location of $a_{\rm{crit}}$ can strongly vary depending on the values of these unconstrained parameters, even though the "statistical orbit" presented in Sec.\ref{sec:orbstab} might give a reasonable estimate.

\subsection{Circumbinary systems}\label{subsec:circum}

Even if the main focus of this study is first and foremost binaries with S-type orbits, we also compiled a database of all known circumbinary (P-type) systems, which we present in Tab.\ref{tab:Ptype} and illustrate in Fig.\ref{fig:circum}. The size of this sample is much more limited, as there are only 31 circumbinary planetary systems known to date, more than 20 times less than in our main S-type database. 
Adding these systems to the 728 S-type binaries thus only marginally changes the global multiplicity fraction, which increases from 16.1 to 16.7$\%$, and from 22.5 to 23.0$\%$ for the $d_b\leq500\,$ domain.

As mentioned in the previous section, four of these circumbinaries also host a more distant stellar companion, and are hierarchical triples for which the planet is both on a P-type and an S-type orbit. Interestingly, 4 out of 31 gives an "S-type" multiplicity fraction of $\sim13\%$, which is roughly comparable to our global S-type multiplicity rate of 16.1$\%$, even if this conclusion suffers from obvious small number statistics.

\begin{figure*}[h!]
\includegraphics{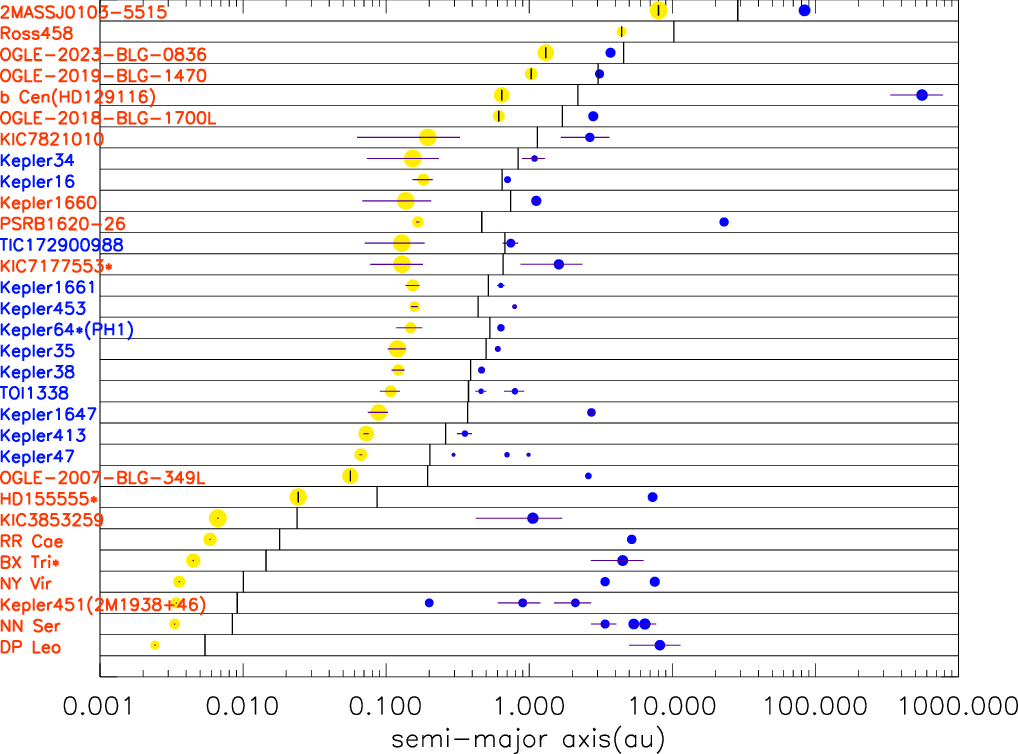}
\caption{{Architecture of the 31 circumbinary planetary systems known to date. Note that we plot here, for both the companion star and the planet(s), the semi-major axis centered on the binary's center of mass}
\label{fig:circum}}
\end{figure*}

\section{Discussion}\label{sec:discus}

In the previous section we have argued that, despite the heterogeneity of the sources our database is built on, some robust statistical results and trends can be derived.
This is notably the case for the deficit of planet-hosting close-in binaries and the reliability of the global multiplicity fraction that we have derived in the $d<500\,$ domain.
Let us here discuss these important issues in more detail.

\subsection{Deficit of close-in binaries}\label{sec:closein}

The lack of close-in companion stars displayed in Fig.\ref{fig:sep} is a feature that has been identified in numerous previous surveys of planet-hosting multiples \citep{wang14,kraus16,mug19,ziegler20}. Even if such a deficit makes sense from a theoretical point of view, because of the expected detrimental effect of close companions on the planet-formation process \citep{theb15}, there is an ongoing debate as to whether or not this observed feature is due to observational biases or not (see Sec.\ref{sec:intro}).
We argue that Fig.\ref{fig:distang}, showing that the lower envelope of the ($\theta_b,d_b$) distribution is not horizontal, convincingly demonstrates that the close-binary deficit cannot be entirely due to biases. This is because the two major biases potentially affecting our sample, which are the adverse selection effect against tight binaries in RV surveys \citep{hirsch21} and the difficulty of finding close companions around planet-hosts in deep imaging surveys \citep{kraus16}, do not depend on the physical separation $\rho_b$ between binary components but on their on-sky \emph{angular} separation $\theta_b$. This would lead to a horizontal lower envelope in a ($\theta_b,d$) graph, which is clearly not the case. One could argue that this non-horizontal lower envelope is due to the fact that the ($\theta_b,d_b$) graph is dominated by RV systems (black dots) at small $d_b$ and transit ones (blue dots) at large $d_b$, and that the different biases affecting these two different populations leads to a jump at the interface between these two distance domains. It is easy to see, however, that the non-horizontal envelope is observed even when only restricting the sample to RV systems  or when only considering transit-systems. In addition, not only is the lower envelope non-horizontal, but it almost follows a diagonal $\rho_b$=cst line, which is what would be expected if the depletion of close binaries was a real physical feature. 
This analysis expands on the argument made by \cite{ziegler20} that the depletion of close-in companions must be real because it was of comparable amplitude for KOIs and TOIs despite of the fact that Kepler targets were located at larger distances.

Of course, we are aware that our sample cannot be fully bias-free, and that some of the lack of close-in binaries is due to adverse selection effects and limits to companion-seeking imaging searches, but it cannot be all of it.
In contrast, it is almost certain that the lack of wide separation binaries in the ($d_b\geq 500\,$pc, $\theta_b\geq$2") upper right-hand corner of Fig.\ref{fig:distang} is a fully artificial feature. It likely results from the fact that two of the largest companion-searching surveys, \cite{mug19} and \cite{michmug24}, which account for roughly 40$\%$ of our sample, were limited to $d_b\leq500-600\,$pc (because of precision limits on Gaia DR2+DR3 parallax and proper motion measurements) and that most objects beyond $500\,$pc come from the surveys by \cite{hirsch17} and \cite{sull23,sull24}, which vetted numerous stellar companion candidates identified in \cite{furlan17}, but only for on-sky separation $\theta_b\leq$2".

\subsection{Multiplicity fraction at $d<500\,$pc}\label{sec:mul500}

A similar argument can be made with respect to the $f_M$ vs. $d_b$ distribution shown in Fig.\ref{fig:distfra}. Indeed, the fact that the two major adverse biases affecting binary statistics become more pronounced with decreasing $\theta_b$ should mathematically lead to smaller $f_M$ values at larger distances in a bias-dominated sample. Granted, such a decrease is observed, but only in the $d_b>500\,$pc region. In the $d_b\lesssim500\,$pc domain, on the contrary, $f_M$ is almost constant, strongly arguing that, even if a bias is present, it is not strong enough to skew global $f_M$ statistics in this domain.

Similarly to the separation distribution, this does not imply that our sample is bias-free, even at $<500$pc. As an example, a 50au physically separated binary is highly likely to be excluded from RV surveys if it is further away than $\sim50$\,pc, corresponding to an on-sky separation $\leq1$". We note, however, that the vast majority of RV-detected systems are relatively close, with $\sim70\%$ of them lying at less than 60\,pc, and that the population of exoplanets is largely dominated by transiting planets beyond this distance (Fig.\ref{fig:methbase}). So that RV systems significantly affect the global value of $f_M(d_b)$ only in a distance domain where the selection bias, albeit real, is relatively limited. 
In addition, the $\sim$1-2" limit for excluding stellar companions is in fact not an absolute rule in RV surveys. When indeed the magnitude difference $\Delta_M$ between the stellar components is higher than $\sim$3-4, signal pollution by the companion becomes small enough to allow correct RV analysis even when the companion is closer than 1" \citep{eggen07}. This is the explanation for most of the planets detected by RV in $\lesssim25\,$au binaries, such as HIP90988, HD72892, HD144899, HD28192, HD199509 or HD145934. And in some rare cases, RV detections are even possible in tight binaries with $\Delta_M<3$ (HD196885, HD87646).

\begin{figure}[h!]
\includegraphics[scale=0.5]{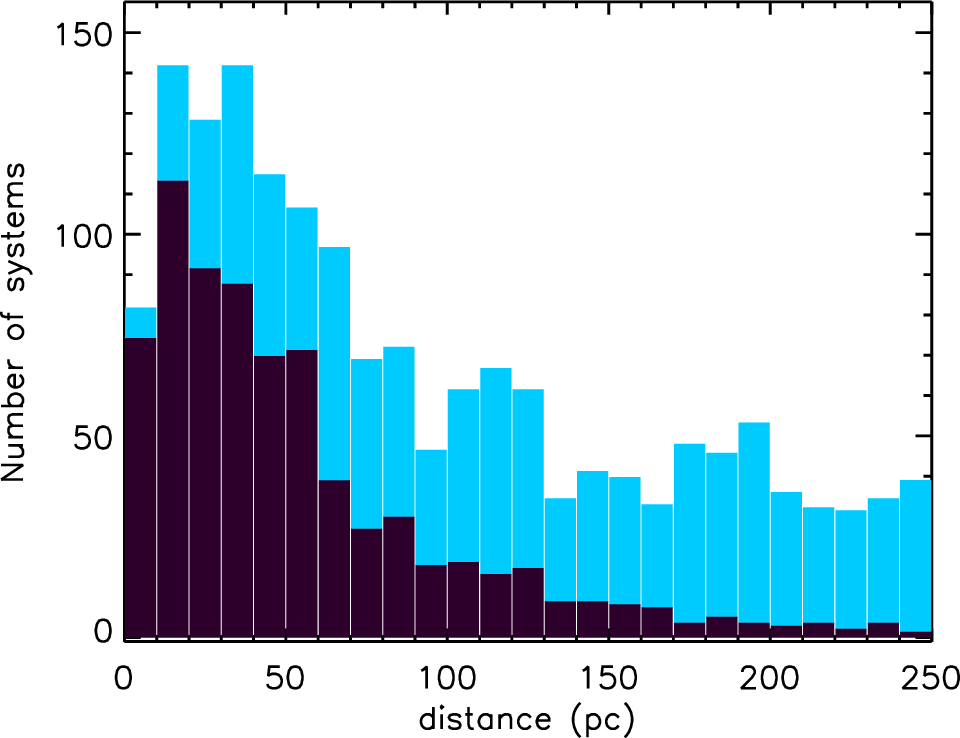}
\caption{{Number of planet-hosting systems as a function of distance for the whole sample of 4504 systems listed in \url{https://exoplanet.eu/}. Blue is for transit detected planets and deep purple for RV planets.  }.
\label{fig:methbase}}
\end{figure}

Having identified the $\lesssim500$pc domain as providing more reliable (if not fully bias-free) statistics, we can use this to derive more accurate results regarding the separation distribution of planet-hosting binaries. Fig.\ref{fig:binfrac} shows the fraction $f_{\rm{M}}(a_b)$ of planet-hosting stars having a stellar companion with a given semi-major axis, for both the total sample and the volume limited $d_b\leq500\,$pc one, as compared to the empirical distribution for field stars derived by \cite{raghavan2010}.
It can be seen that, while the $f_{\rm{M}}(a_b)$ curve for the whole sample always stays below that of field stars, the distribution of $f_{\rm{M}}(a_b)$ in the $d_b\leq500\,$pc domain becomes comparable to that of field stars for $a_b\gtrsim$300-500\,au\footnote{One could argue that the match beyond $a_b\gtrsim$300-500\,au is not perfect, with an excess of companions in the 1000-5000\,au range and a deficit of companions beyond $\sim10000\,$au, but these features are actually also observed for field stars: the empirical law derived by \cite{raghavan2010} indeed underestimates $f_M(a_b)$ for $1000\lesssim a_b \lesssim 5000\,$au and overestimates it beyond $\sim10000\,$au (see Fig.13 of that paper)}. This is a clear indication of the completeness of the search for stellar companions to planet-hosts, at least for wide companions. It also strongly suggests that the deficit of planet-hosting close-in binaries could extend to separations of $\sim$300-500\,au.

\begin{figure}[h!]
\includegraphics[scale=0.5]{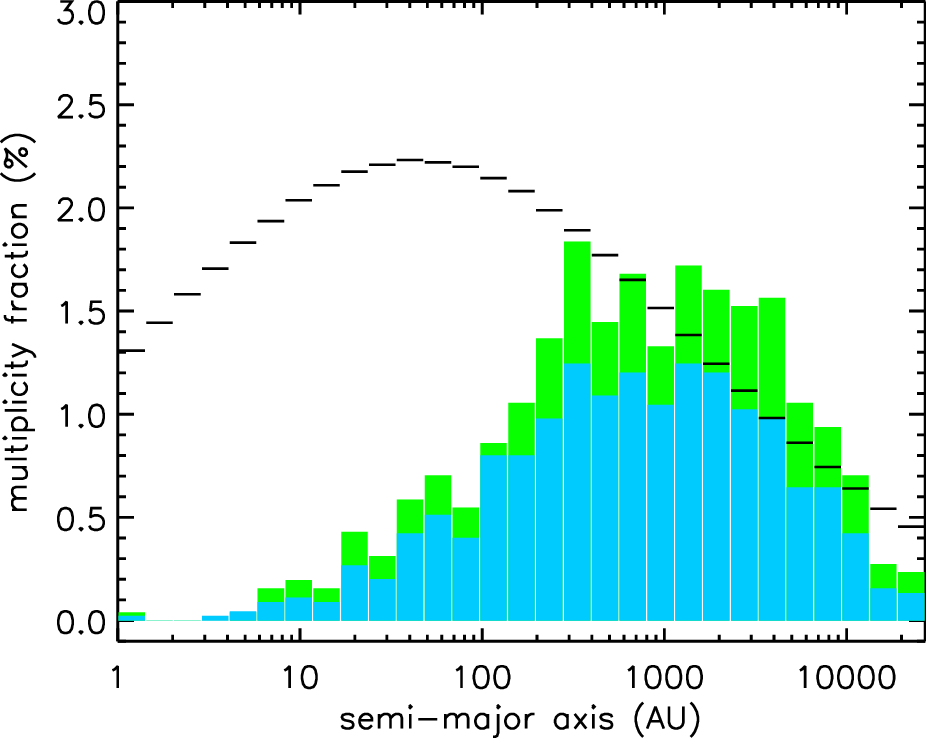}
\caption{{Fraction $f_{\rm{M}}(a_b)$ of planet-hosting stars having a stellar companion in a given semi-major axis range. The blue histogram correspond to the whole population of planet-hosting binaries, while the green one is computed for systems at a distance of less than 500pc. Values of $f_{\rm{M}}(a_b)$ are calculated by dividing the number of systems in each $a_b$ bin by the total number (or the total number at $d_b\leq500\,$au) of planet-hosting stars found in the \url{https://exoplanet.eu/} database. The black segments represent the multiplicity fraction distribution for the canonical binary distribution of \cite{raghavan2010} (note that, contrary to Fig.\ref{fig:sep}, the \cite{raghavan2010} distribution is here not normalized, allowing direct comparison to our database)}.
\label{fig:binfrac}}
\end{figure}

\subsection{Accounting for completeness issues}\label{bias}

\begin{figure}[h!]
\includegraphics[scale=0.5]{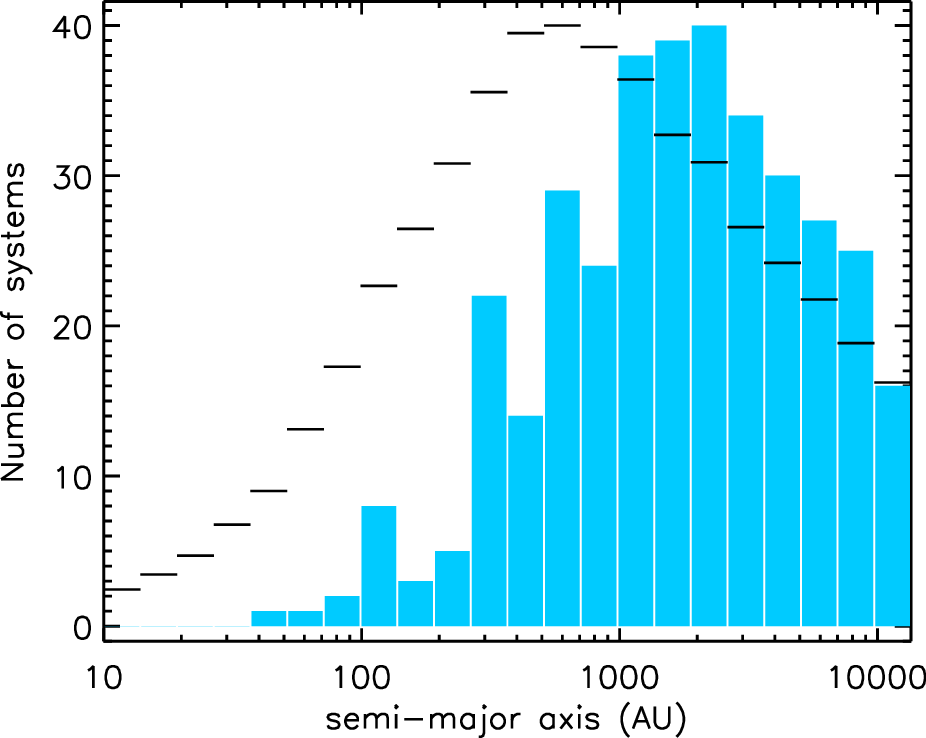}
\caption{{Histogramm (in blue) of the $a_b$ distribution for the subsample of Gaia-derived planet-hosting binaries obtained from \cite{mug19}, \cite{michmug21} and \cite{michmug24}. We also present the normalized $a_b$ distribution for a Monte-Carlo generated population of field binaries following the canonical distribution of \cite{raghavan2010}} and the Gaia detection limit presented in \cite{mug22} (see main text for details).
\label{fig:bias}}
\end{figure}

Our main result that the deficit of planet-hosting companions at $\lesssim$ 500\,au separations cannot be fully explained by biases is essentially based on a visual analysis of the angle versus distance (Fig.\ref{fig:distang}) and $f_M$ versus distance (Fig.\ref{fig:distfra}) plots. Another, and in principle more reliable (and more direct) way to derive this result would have been to accurately quantify all the observational biases (completeness and detection limits) affecting our database in order to debias it and compare it to the characteristics of field binaries.
Unfortunately, the extreme diversity and heterogeneity of observational sources that have been used in constructing the database prevents us from carrying out this procedure. Even when only considering surveys that have provided more than 10 objects, there are still 21 different observational sources left, each with its own specific detection limits and target selection policy. There are, in addition, the complex (and frequent) cases for which the companion stars have been visually detected in an AO or Speckle imaging survey but have been confirmed as bound companions in one or several later studies, whose limitations and target selection policy have also to be factored into the bias estimate.

We do, however, make an attempt at partially assessing the level of bias by focusing on the several surveys that have mined the Gaia catalogue, and which have relatively coherent detection limits. This is especially true for the series of surveys that Mugrauer and collaborators published since 2019, which make up almost $40\%$ of our database, for which we assume the empirical detection limit $\Delta M(\theta)$ derived by \cite{mug22}, where $\Delta M$ is the magnitude difference between the binary components and $\theta$ is their angular separation \citep[see Fig.3 of][]{michmug24}. We then use a Monte Carlo (MC) scheme to generate a sample of $10^{5}$ binaries following the canonical distribution for field binaries by \cite{raghavan2010} and the target distance distribution displayed in Fig.1 of \cite{michmug24}, and retain only those satisfying the aforementioned detection limit. Fig.\ref{fig:bias} presents the obtained separation distribution for this sample and compare it to the distribution for the "MG" subsample of companions obtained by \cite{mug19}, \cite{michmug21} and \cite{michmug24}. It can be  seen that, while the two distribution profiles do approximately match in the $\gtrsim 1000\,$au domain, there is a significant lack of short separation binaries for the MG subsample as compared to what would be expected for a observationally biased field binary population. 
The MC population peaks at a separation of $\sim500\,$au while the MG subsample peaks around 2000\,au. Using a bootstrap resampling procedure on the MG subsample we find that its mean separation is comprised between 1645 and 2432\,au at a confidence level of $3\sigma$ on the logarithmic distribution of separations. It is thus statistically different from that of the MC-generated population, whose average separation (calculated from the logarithmic distribution) is 727\,au. Furthermore, there are only 4 planet-hosting binaries with $a_b\leq 100\,$au in our MG subsample, when we should have had $\sim56$ such systems for a canonical \cite{raghavan2010} distribution (Fig.\ref{fig:bias}).

This confirms that, for this important subsample of the database, the population of planet-hosting binaries physically differs from that of field binaries. However, it is also clear that some observational bias is present, as indicated by the difference between the MC distribution profile and that of the unbiased \cite{raghavan2010} binary population displayed in Fig.\ref{fig:sep}.

\subsection{Stability and dynamical environment}\label{sec:dynenv}

The detailed study of individual systems is not the purpose of the present work. We presented however in Fig.\ref{fig:binstruc} a global view of the dynamical architecture of the tightest binaries in our sample, showing in particular the location of the planet(s) with respect to the empirical orbital stability limit $a_{\rm{crit}}$ \citep{holwig99}. As discussed in Sec.\ref{sec:dynenv}, all but 2 planets are in the "safe" $a_b\leq a_{\rm{crit}}$ region, and even these two could in fact be on a stable orbit. It is, however, important to stress that the $a_b\leq a_{\rm{crit}}$ criteria should only be taken as a first approximation and that precise orbital stability can only be assessed by means of thorough numerical exploration.

Going beyond the sole stability criteria, we note that a handful of planets, all giants, such as HIP90988b, HD41004Ab, TOI1736c, HD196885b, $\gamma$ Ceph b, GJ9714b or HD217958b, orbit at locations that are below $a_{\rm{crit}}$ but relatively close to it (Fig.\ref{fig:binstruc}). So that, even if their orbits are stable, their dynamical environment is strongly perturbed by the companion star. Investigating planet formation processes in binaries goes far beyond the scope of the present paper, but it is highly likely that the presence of a companion star must have strongly perturbed the formation of these planets. As briefly evoked in Sec.\ref{sec:intro}, the formation stage that is probably most affected by binary perturbations is the planetesimal accretion phase, as it requires a dynamically quiet environment with low enough impact velocities. Quantifying this detrimental effect is, however, a (very) arduous task, as it depends on the complex interplay between several mechanisms, such as binary secular perturbations, gas friction, disc self gravity or collisional output prescriptions \citep{marz19}. Several analytical and numerical studies have nevertheless investigated this issue, but have often reached conflicting conclusions regarding the detrimental effect of binarity \citep[see, for instance,][]{theb08, paard08, xie09, xie11, frag11, rafi13, sils21}. Explaining the presence of giant planets in the dynamically highly-perturbed environment close to $a_{\rm{crit}}$ thus remains an open issue. 
Furthermore, planet-formation in binaries has so far only been investigated in the context of the "classical" incremental core-accretion scenario, in which km-sized planetesimals slowly form from the local mutual sticking of smaller grains, and in turn grow by mutual gravity during two-body impacts \citep[e.g.,][]{liss93}. To our knowledge, no study has attempted to explore the new paradigm of planet-formation that has emerged in recent years, and the role of, for instance, streaming instability \citep[e.g.,][]{johan15}, pebble accretion \citep{johan17} or snow-line induced pressure bumps \citep{izi22}, in the context of binarity. 
We believe that our database could be of great use for future studies investigating these issues. It could be exploited to test these mechanisms observationally, and in particular predictions they make regarding the correlations between binary separations and some other key observables, in particular planetary masses and planetary locations.

\section{Summary and Conclusion}\label{sec:conclu}

We have compiled the most extensive database of planet-hosting multiple systems (mainly binaries) ever assembled. This database currently lists the main characteristics (binary orbit or projected separation, distance, stellar masses, number of planets, stability limits for planets, etc.) of 728 S-type (circum-primary or circum-secondary) and 31 P-type (circumbinary) systems. It is made available to the community in an online version that will be regularly updated. 

The heterogeneity of the sources this database is built on, and the different observational biases that affects them, is a clear challenge for deriving precise statistical correlations. Some important qualitative trends and results can, however, be derived. We confirm that the raw separation distribution of planet-hosting binaries shows a pronounced deficit of close-in companions. Plotting the
distribution of binary \emph{angular} separations as a function of system distance $d_b$, we observe a non-horizontal lower envelope that indicates that this deficit of planet-hosting close-in binaries cannot be entirely due to adverse observational biases. We also show that the multiplicity fraction $f_M$ among planet-hosting stars does not vary with system distance up to $d_b\sim500\,$pc, indicating that the subsample of $d_b\lesssim500\,$pc planet-hosting binaries is not bias-dominated (without being bias-free). For this subsample we find $f_M=22.5\%$, roughly half the multiplicity fraction of solar-type field stars, which gives a first order estimate of the detrimental effect binarity has on planet-formation. For the same $d_b\lesssim500\,$pc subsample, we find that the distribution of multiplicity fraction $f_M(a_b)$ as a function of binary separation closely matches that of field stars beyond $\sim$300-500\,au, indicating that the search for stellar companions to planet-hosting stars is mostly complete at these separations. It also strongly suggests that the detrimental effect of binarity on planet-formation can be felt up to separations of $\sim$500\,au.

\bibliographystyle{aa}
\bibliography{main}

\appendix

\section{circumbinary database}

\begin{table*}[]
\begin{center}
\begin{tabular}{c c c c c c c c c c c c c c}
\hline
Name  & Alt. Name &$M_1$ & $M_2$ & $d$ & method & $a_b$ (or $\rho_b$) & $e_b$ & $n_{\rm{Pl}}$ & $a_{\rm{Pl}}$ & $e_{\rm{Pl}}$ & $m_{\rm{Pl}}$ & $a_{\rm{crit}}$ & $i_{\rm{b}}$\\
 & (if any) & ($M_{\oplus}$) & ($M_{\oplus}$) & (pc) & & (au) & & & (au) & & ($M_{\rm{Jup}}$) & (au) & (deg.)\\
\hline\hline
DP Leo & X & 0.690 & 0.090 & 305.0 & 3 & 0.0027 & 0. & 1 & 8.19 & 0.390 & 6.05 & 0.0054 & - \\
NN Ser & X & 0.535 & 0.111 & 521.0 & 3 & 0.0039 & 0. & 3 & 3.39 & 0.20 & 2.28 & 0.0084 & - \\
Kepler451 & 2M1938+46 & 0.480 & 0.120 & 400.0 & 3 & 0.0041 & 0. & 3 & 0.2 & 0. & 1.76 & 0.0091 & - \\
NY Vir & X & 0.460 & 0.140 & 595.0 & 3 & 0.0044 & 0. & 2 & 3.39 & 0. & 2.78 & 0.0101 & - \\
BX Tri* & X & 0.510 & 0.210 & 59.0 & 3 & 0.0060 & 0. & 1 & 4.5 & 0.4 & 7.5 & 0.0142 & - \\
........\\
\hline
\end{tabular}
\end{center}
\caption{Database of planet-hosting "P-type" (circumbinary) binaries, sorted by increasing binary separation. This is just
an excerpt of the full table that is available, in a machine-readable form, at this url: \url{https://exoplanet.eu/planets_binary_circum/} }
\label{tab:Ptype}
\end{table*}

\section{Notes and references on S-type planet-hosting binaries}

\begin{table*}[]
\begin{center}
\begin{tabular}{c l }
\hline\hline
Gamma Ceph &  Post-main-sequence star (sub-giant). Astrometry analysis \citep{bene18} shows that there is a 70 degrees mutual\\
 & inclination between the binary and planetary orbits and that the planet probably has a much higher mass than previously\\
 & expected ($9M_{\rm{Jup}}$). Binary orbit last revised by \cite{mug22}. Host star visible to the naked eye.\\
HD196885 & Binary by \cite{chauv06}, parameters last refined by \cite{chauv23}. Note that \cite{barba23} find a\\
 & much shorter separation (13au) for the binary\\
16 Cyg B* & Binarity in \cite{ragha06}, parameters updated by \cite{michmug24}. Hierarchical triple: the "companion"\\
 & consists of two stars (16 Cyg A\&C) separated by 73au. These two stars' mass are summed up in the database to create a\\
 & "virtual" companion.\\
KIC 7177553* & Planet and stellar components by \cite{conr14,lehm16,borko16}. Quadruple system:\\
 & the central "star" is a tight binary and the companion "star" is also a tight binary (period 16d). Very rare case of a system\\
& with a planet that is both on a P-type and a S-type orbit\\
Kepler1433& Binarity hinted at by \cite{furlan17}, confirmed by \cite{sull23} as bound\\
(KOI3120)  & \\
........ & ........\\
\hline
\end{tabular}
\end{center}
\caption{Notes and references for each individual system of the S-type database. This is only a highlight of the full notes that can be accessed at this url: \url{https://exoplanet.eu/planets_binary_notes/} }
\label{tab:notes}
\end{table*}

\section{Notes and references on P-type (circumbinary) planet-hosting binaries}

\begin{table*}[]
\begin{center}
\begin{tabular}{c l }
\hline\hline
DP Leo & planet detected by timing by \cite{qian10} / Post-common envelope cataclysmic binary\\
NN Ser & First planets hinted at in \cite{beuer10}, still debated but claimed as probably confirmed by \cite{ozdo23}. \\
 & Planets' characteristics not fully constrained yet / Post-common envelope cataclysmic binary\\
Kepler-64* & Planet by \cite{kostov13} and \cite{schwa13} / Quadruple system with another (60au separated) binary at 1000au\\
(PH1) & from the central one. Rare case of  planet on both a P-type and an S-type orbit\\
b Cen & Planet by \cite{jans21} , whose eccentricity is very loosely constrained (0.4 $\pm$0.4)/ stellar companion by\\
(HD129116) & \cite{rizzu13} / Stellar masses derived by Squiciarrini (2025, personal communication) \\
........ & ........\\
\hline
\end{tabular}
\end{center}
\caption{Notes and references for each individual system of the P-type (circumbinary) database. This is only a highlight of the full notes that can be accessed at this url: \url{https://exoplanet.eu/planets_binary_notescirc/} }
\label{tab:notescirc}
\end{table*}



\end{document}